\documentstyle[floats,multicol,aps,epsf,prb]{revtex}

\newcommand \be  {\begin{equation}}
\newcommand \ee {\end{equation}}
\newcommand \bea {\begin{eqnarray} }
\newcommand \eea {\end{eqnarray}}

\begin{document}
\draft
\twocolumn[\hsize\textwidth\columnwidth\hsize\csname @twocolumnfalse\endcsname
\title{Level correlations in disordered superconducting grains}
\author{Revaz Ramazashvili}
\address{
Loomis Laboratory, University of Illinois at Urbana-Champaign, \\ 
Urbana, IL 61801-3080, USA. }
\maketitle
\date{\today}
\maketitle
\begin{abstract}
I study the quasiparticle level correlations in a grain of a weakly disordered
d-wave superconductor, and show that, in a wide intermediate energy range, they
are characterized by a novel type of universal behavior.
\end{abstract}
\vskip 0.2 truein
\pacs{PACS numbers: 73.23.-b}
\vskip2pc]

Level correlations represent a fundamental property of an electron system. They
reflect sensitivity of the quasiparticle spectrum to disorder and boundary conditions,
and distinguish between extended and localized states. Level correlations in
metal grains, \cite{ge,e,as,sa} mesoscopic wires, \cite{af} quantum dots,
\cite{az2} SNS junctions, \cite{asts} and superconducting vortex cores \cite{skf,bcsz,kl,i}
have recently been a subject of much active study. Most simply, level correlations
are quantified by the mean square deviation \( \langle \delta N^{2}_{E}\rangle  \)
of the number \( N_{E} \) of levels in an energy interval of width \( E \):

\[
\langle \delta N^{2}_{E}\rangle \equiv \langle \left[ N_{E}-\langle N_{E}\rangle \right] ^{2}\rangle ,\]
 where the angular brackets denote disorder average. 

In disordered systems, the energy levels of localized states are uncorrelated,
and \( \langle \delta N^{2}_{E}\rangle \sim \langle N_{E}\rangle  \). By contrast,
extended states are strongly correlated, \cite{e,as} which is expressed in
\( \langle \delta N^{2}_{E}\rangle  \) scaling as only a logarithm of \( \langle N_{E}\rangle  \):

\begin{equation}
\label{eq:answer}
\langle \delta N^{2}_{E}\rangle \approx \frac{2C}{\pi ^{2}}\ln \langle N_{E}\rangle \ll \langle N_{E}\rangle ,
\end{equation}
 where \( C \) is the number of the quasiparticle diffusion modes. \cite{remark1}
Formula (\ref{eq:answer}) not only measures the level correlations, but also
shows that, in a disordered system, the level number variance is driven by diffusion
modes. Formula (\ref{eq:answer}) holds for \( E \) smaller than the Thouless
energy \( E_{c}\equiv D/L^{2} \), where \( D \) is the diffusion constant,
and \( L \) is the system size. At these energies, the level correlations are
remarkably universal, with the constant \( C \) being defined solely by the
fundamental symmetries of the system. The low energy universality classes of
disordered systems have been classified based on the symmetry arguments. \cite{m,az1}

In this article, I study quasiparticle level correlations in a disordered grain
of a d-wave superconductor in the presence of both the time reversal (\( T \))
and the spin rotation invariance (\( S \)). I show that, in a wide intermediate
energy range, the level correlations have the universal form (\ref{eq:answer}),
yet are different from those in a grain of a metal or of an s-wave superconductor
(also invariant under \( T \) and \( S \)) -- as well as different from the
level correlations in any of the previously charted \cite{m,az1} low energy
universality classes. The main result is encapsulated in the number \( C \)
of the diffusion modes. In a grain of normal metal (or an s-wave superconductor)
with both the time reversal and the spin rotation symmetries present, \( C=4 \),
since both the quasiparticle charge and the three components of the quasiparticle
spin are conserved and propagate diffusively. By contrast, I show that, in a
d-wave superconducting grain, the charge diffusion mode disappears, which leads
to \( C=3 \), despite the very same set of fundamental symmetries. This is
a novel universal type of level correlations. \emph{}

The reason behind this result is that the impurity scattering leads to the exchange
of charge between the quasiparticle subsystem and the condensate. However, this
process is sensitive to the momentum anisotropy of the gap, and its rate vanishes
for an ideally isotropic gap, as noticed long ago \cite{t} in the context of
the branch imbalance relaxation in NS junctions. On the contrary, in a d-wave
superconductor, such a charge relaxation occurs at a time scale of order the
impurity scattering time. As a result, compared with a normal metal (or an s-wave
superconductor), the quasiparticle charge diffusion mode is missing in a d-wave
superconductor, which reduces the constant \( C \) from four to three. This
reduction is a robust many-body effect of the anisotropic pairing symmetry. 

The plan of the paper is as follows. First, I show that, in an s-wave superconducting
grain, the level correlations are essentially the same as in the normal state.
Then I show how the gap anisotropy eliminates quasiparticle charge diffusion
in a d-wave superconductor, and outline the corresponding microscopic calculation.
Finally, I describe the applicability range of the result and its relation to
the previous findings, and discuss the possible further developments. 

\textbf{S-wave superconductor.} Consider an s-wave superconducting grain. In
the approximation of a spatially uniform gap, superconductivity can be described
as pairing of the exact time reversed eigenstates of the underlying metal.\cite{dg}
Thus, the exact quasiparticle energies \( E_{n} \) in the superconducting state
can be expressed via the exact quasiparticle energies \( \varepsilon _{n} \)
in the normal state as per \( E_{n}=\sqrt{\varepsilon ^{2}_{n}+\Delta ^{2}} \).
Therefore, the exact density of states \( \nu _{S}(E)\equiv \sum _{n}\delta [E-\sqrt{\varepsilon ^{2}_{n}+\Delta ^{2}}] \)
in the superconducting state is simply related to the exact density of states
\( \nu _{N}(\varepsilon )\equiv \sum _{n}\delta [\varepsilon -\varepsilon _{n}] \)
in the normal state:  

\begin{eqnarray*}
\nu _{S}(E)
 &=&
\frac{E}{\sqrt{E^{2}-\Delta ^{2}}}\sum _{n}\delta [\sqrt{E^{2}-\Delta ^{2}}-\varepsilon _{n}]  \\
 &=&
\frac{E}{\sqrt{E^{2}-\Delta ^{2}}}\nu _{N}(\sqrt{E^{2}-\Delta ^{2}}).
\end{eqnarray*}

As a result, the density of states (DoS) dimensionless autocorrelation function
\( R^{S}_{2}(E,E')\equiv \frac{\langle \nu _{S}(E)\nu _{S}(E')\rangle }{\langle \nu _{S}(E)\rangle \langle \nu _{S}(E')\rangle }-1 \)
in the superconducting state can be expressed through the DoS autocorrelation
function in the normal state \( R^{N}_{2}(\varepsilon ,\varepsilon ')\equiv \frac{\langle \nu _{N}(\varepsilon )\nu _{N}(\varepsilon ')\rangle }{\langle \nu _{N}(\varepsilon )\rangle \langle \nu _{N}(\varepsilon ')\rangle }-1 \)
in the form

\[
R^{S}_{2}(E,E')=R^{N}_{2}(\sqrt{E^{2}-\Delta ^{2}},\sqrt{(E')^{2}-\Delta ^{2}}).\]

From this simple argument, it follows that the level correlations in an s-wave
superconductor are identical (up to the change of variables) to those in the
underlying normal metal and that, therefore, the diffusion modes in the two
systems are the same. 

Before moving further, it is instructive to classify the quasiparticle diffusion
modes more precisely. As a two-particle process, diffusion amounts to a coherent
propagation of a particle and a hole, carrying spin-\( \frac{1}{2} \) each.
These two spins can add to form a singlet, which corresponds to the charge degree
of freedom -- or a triplet, which corresponds to the three spin degrees of freedom.
Both in s- and in d-wave superconductor, the condensate carries no spin, and
thus a quasiparticle cannot exchange spin with the condensate. As a result,
in a disordered spin-singlet superconductor, the quasiparticle spin does propagate
diffusively, as it was recently re-emphasized. \cite{sf} By contrast, the situation
with the quasiparticle charge is more delicate. Perhaps the simplest way to
observe this difference is by inspecting the Bogolyubov quasiparticle creation
operator:

\[
\gamma ^{+}_{p\uparrow }=u_{p}c^{+}_{p\uparrow }+v_{p}c_{-p\downarrow } ,\; u^{2}_{p}=\frac{1}{2}[1-\frac{\varepsilon _{p}}{\sqrt{\varepsilon ^{2}_{p}+\Delta ^{2}_{p}}}] ,\; u^{2}_{p}+v^{2}_{p}=1.\]
 Impurity scattering is elastic, i.e. it conserves the quasiparticle energy
\( E_{p}=\sqrt{\varepsilon ^{2}_{p}+\Delta ^{2}_{p}} \) . In an s-wave superconductor
with the uniform gap, \( \Delta _{p} \) is a constant and, in the absence of
the Andreev scattering that turns \( \varepsilon _{p} \) into \( -\varepsilon _{p} \),
the energy conservation implies conservation of \( u_{p} \) and \( v_{p} \)
. This means conservation of the particle-hole content of a quasiparticle and
leads to the effective charge conservation -- even though a Bogolyubov quasiparticle,
being a superposition of a particle and a hole, does not have a well defined
charge quantum number. The same conclusion can be reached by considering directly
the expectation value of the quasiparticle charge \( Q_{p} \) : 

\[
Q_{p}=u^{2}_{p}(+1)+v^{2}_{p}(-1)=\frac{\varepsilon _{p}}{\sqrt{\varepsilon ^{2}_{p}+\Delta ^{2}_{p}}}.\]
In an s-wave superconductor, and in the absence of the Andreev processes, \( Q_{p} \)
is conserved by the impurity scattering, which leads to the charge diffusion
pole. 

\textbf{D-wave superconductor.} By contrast, in a d-wave superconductor, the
gap \( \Delta _{p} \) has strong momentum dependence. Hence, even in the absence
of the Andreev scattering processes, \( Q_{p} \) is not conserved by impurity
scattering. In other words, in a d-wave superconductor, elastic scattering does
not conserve the moduli of the Bogolyubov factors \( u_{p} \) and \( v_{p} \)
, thus changing the particle-hole content of a quasiparticle. Physically, this
means that the impurity scattering leads to the exchange of charge between the
quasiparticle subsystem and the condensate at the time scale of order the impurity
scattering time -- and thus to the absence of the charge diffusion pole.

The same conclusion can be reached in a different and more formal way, by using
the Ward identities, which are a consequence of the symmetries of the system.
\cite{s} In the Nambu notations, the BCS Hamiltonian of a d-wave superconductor
reads

\begin{eqnarray*}
H=\int dr\Psi ^{\dagger }\left[ \tau _{3}\varepsilon (\vec{p}-\frac{e}{c}\vec{A}\tau _{3})+\tau _{3}e\phi -\mu B+\tau _{3}u\right] \Psi + &  & \\
+\int dRdr\Psi ^{\dagger }(R+\frac{r}{2})\tau _{1}\Delta (R,r)\Psi (R-\frac{r}{2}). &  & 
\end{eqnarray*}

Here \( \Psi ^{\dagger }\equiv (\psi _{\uparrow }^{\dagger },\psi _{\downarrow }) \)
is the Nambu spinor, \( \tau _{i} \) are the Nambu matrices, \( R \) denotes
the center of mass coordinate of a Cooper pair and \( r \) denotes the relative
coordinate. The external fields are the vector potential \( \vec{A} \), the
electric potential \( \phi  \), the Zeeman field \( B \) in the \( \hat{z} \)
direction, and \( u \) is the impurity potential. The pair field \( \Delta (R,r) \)
has been chosen real for the sake of simplicity, and assumed to have the d-wave
angular dependence. 

The Hamiltonian above respects the gauge symmetry \( \Psi \rightarrow U_{g}\Psi =\exp \left[ i\tau _{3}\frac{e}{\hbar c}\chi _{g}\right] \Psi  \)
(to be accompanied by the appropriate change of the potentials and of the gap
function), and the \( \hat{z} \) axis spin rotations symmetry \( \Psi \rightarrow U_{s}\Psi =\exp \left[ i\frac{e}{\hbar c}\chi _{s}\right] \Psi  \)
(to be accompanied by the change of \( B \)). \cite{remark2} The sought identities
can be obtained in the usual way, \cite{agd} by identifying the variation of
the Green function under an infinitesimal transformation \( G\rightarrow UGU^{+} \)
with the first-order perturbative correction. In the absence of the external
fields \( \vec{A} \), \( B \) and \( \phi  \), and assuming spatially uniform
gap (\( \Delta (R,r)=\Delta (r) \)), the two symmetries imply two (\( Q=0,\Omega \equiv \varepsilon -\varepsilon '\rightarrow 0 \))
Ward identities for the disorder-averaged Green functions \( G_{R,A}^{-1}(\varepsilon ,p)=\varepsilon -\Sigma _{R,A}(\varepsilon )-\tau _{1}\Delta _{p}-\tau _{3}\varepsilon _{p}\, ,\, \Sigma _{R}(\varepsilon )=\Sigma _{A}^{\ast }(\varepsilon ) \)
and the vertex renormalizations \( \langle \tau _{0}\rangle _{RA} \) , \( \langle \tau _{3}\rangle _{RA} \)
and \( \langle \Delta _{p}\tau _{2}\rangle _{RA} \) : 

\begin{equation}
\label{eq:WI-charge}
-2i\tau _{3}\Sigma _{R}''(\varepsilon )=(\varepsilon -\varepsilon ')\langle \tau _{3}\rangle _{RA}+2i\langle \Delta _{p}\tau _{2}\rangle _{RA},
\end{equation}

\begin{equation}
\label{eq:WI-spin}
-2i\Sigma _{R}''(\varepsilon )=(\varepsilon -\varepsilon ')\langle \tau _{0}\rangle _{RA}.
\end{equation}

\begin{figure}
\centering\epsfxsize=3.0 truein
\epsfbox{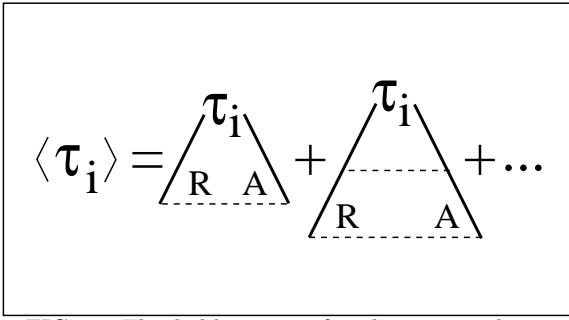}
\caption[]
{The ladder series for the renormalization of the vertices \protect\( \langle \tau _{i}\rangle \protect \). 
}
\label{fig:vertex}
\end{figure}

The Nambu matrix \( \tau _{0} \) corresponds to the \( \hat{z} \)-component
of the quasiparticle spin, and the vertex renermalization \( \langle \tau _{0}\rangle _{RA} \)
describes its propagation. Similarly, \( \tau _{3} \) is the quasiparticle
charge operator, whose propagation is accounted for by \( \langle \tau _{3}\rangle _{RA} \).
The subscript \( RA \) means that the vertex joins a retarded and an advanced
Green function at energies \( \varepsilon  \) and \( \varepsilon ' \). \cite{remark3}
For the illustration purposes, the leading approximation for the vertex corrections
(the ladder series) is shown in Fig. \ref{fig:vertex}.

Direct inspection of the ladder series shows that \( \langle \Delta _{p}\tau _{2}\rangle _{RA}\propto \langle \tau _{3}\rangle _{RA} \),
which leads to the conclusion that the vertex \( \langle \tau _{3}\rangle _{RA} \)
remains finite as \( (\varepsilon -\varepsilon ')\rightarrow 0 \). Thus, the
quasiparticle charge does not propagate diffusively. At the same time, as seen
from Eq. \ref{eq:WI-spin}, the vertex \( \langle \tau _{0}\rangle _{RA} \),
corresponding to the \( z \)-component of the spin, acquires a diffusion pole: 

\[
\langle \tau _{0}\rangle _{RA}=\frac{2\Sigma _{R}''(\varepsilon )}{i(\varepsilon -\varepsilon ')}.\]
 Notice that, in a metal, the charge diffusion mode re-appears, as seen by sending
\( \Delta _{p} \) to zero in Eq. (\ref{eq:WI-charge}).

\textbf{The calculation.} The microscopic calculation of the level correlations
amounts to finding the DoS autocorrelation function \( K(\varepsilon ,\varepsilon ')\equiv \langle \nu (\varepsilon )\nu (\varepsilon ')\rangle -\langle \nu (\varepsilon )\rangle \langle \nu (\varepsilon ')\rangle  \).
The mean square deviation of the number of levels in an energy interval of width
\( E \) is then given simply by  
\[
\langle \delta N^{2}_{E}\rangle =\int _{E}d\varepsilon d\varepsilon 'K(\varepsilon ,\varepsilon ').\]

\begin{figure}
\centering\epsfxsize=3.0 truein
\epsfbox{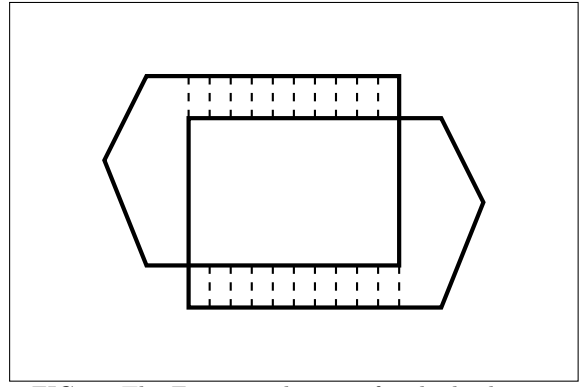}
\caption[]
{The Feynman diagram for the leading term in the DoS autocorrelation function.
}
\label{fig:DoS}
\end{figure}

The calculation of \( K(\varepsilon ,\varepsilon ') \) amounts to evaluating
the Feynman diagram on Fig. \ref{fig:DoS}, as done in \cite{as}, and has to
be performed in the four-spinor Balian-Werthamer space. The technical difference
with regards to a metal being that now the Green functions reside in the matrix
space, whereas the impurity ladder resides in the space of direct products of
the two matrices. The calculation for nodal quasiparticles in a d-wave superconductor
leads to Eq. \ref{eq:answer}, with \( C=3 \), and its details will be published
elsewhere. \cite{remark4}

\textbf{The validity range.} The applicability range of this calculation is
set by the possibility to treat the gap as spatially uniform. \cite{remark5}
Hence the energy interval \( E \) should be much wider than the fluctuations
of the gap. The latter scale is set by the perturbation of the gap due to the
impurity potential. Neglecting the Coulomb interaction, the perturbation of
the gap due to a single impurity has the form \cite{fe}

\[
\delta \Delta (R)\sim \Delta \left[ uN(0)\right] \left[ \lambda N(0)\right] F(R),\]
 where \( \Delta  \) is the value of the gap far from the impurity, \( [uN(0)] \)
is the dimensionless strength of the impurity potential, \( [\lambda N(0)] \)
is the dimensionless BCS coupling constant, and \( F(R) \) is a function decaying
to zero at the lengthscale of order the coherence length. Scattering off \( \delta \Delta (R) \)
is the Andreev reflection off the inhomogeneities of the gap, with the rate
\( \tau ^{-1}_{A}\sim \tau ^{-1}[\Delta /\epsilon _{F}]^{2}[\lambda N(0)]^{2}\ll \tau ^{-1} \),
where \( \tau ^{-1} \) is the impurity scattering rate. Thus, the main result
of this paper holds for \( E \) much greater than \( \tau ^{-1}_{A} \), and
for the levels separated from the Fermi energy by more than \( \tau ^{-1}_{A} \).
At the same time, \( E \) must be much smaller than the Thouless energy \( E_{c}=D/L^{2} \),
\cite{as} which requires \( \tau ^{-1}_{A}\ll D/L^{2} \). The latter inequality
bounds the grain size by 

\[
L\ll \frac{k_{F}l\xi }{[\lambda N(0)]},\]
which is much greater than the coherence length. Note that, for nodal quasiparticles
in a d-wave superconductor, both \( l \) and \( \xi  \) are functions of the
quasiparticle energy \( \varepsilon  \) and scale as \( l(\varepsilon )\sim l\Delta /\varepsilon  \),
which further increases the upper bound on the grain size. 

Before closing, it is instructive to put the main result of this work, Eq. (\ref{eq:answer})
with \( C=3 \), in context. Equation (\ref{eq:answer}) (with different values
of \( C \)) is commonly associated with the random matrix theory (RMT), \cite{m}
which furnishes very general and powerful phenomenological framework for treating
random systems, and allows a symmetry classification \cite{m,az1} of possible
universality classes. However, a crucial underlying assumption of the RMT is
that \emph{all} the matrix elements of the Hamiltonian (including the matrix
elements of the gap) be random and drawn independently from a broad distribution.
This requirement automatically rules out the possibility to distinguish superconductors
of different pairing symmetry. 

By contrast, the present work studies the intermediate energy limit when the
matrix elements of the gap may be treated as completely non-random, the randomness
being restricted to the diagonal of the Bogolyubov-de Gennes Hamiltonian. In
this limit, the Hamiltonian is only a ``partly random'' matrix, with the matrix
elements of the gap fixed by the pairing symmetry. As shown above, the level
correlations in such a ``partly random'' ensemble are sensitive to the momentum
anisotropy of the pairing and are qualitatively different e.g. for s- and d-wave
superconducting grains. 

More importantly, this work shows that, in anisotropic superconductors, the
condensate assumes the role of a ``charge reservoir'' coupled to the quasiparticle
subsystem, and affects the level correlations. It would be interesting to study
this problem further by explicitly including the dynamics of the condensate,
especially in view of the cuprate superconductors as an obvious experimental
object. 

Discussions with E. Abrahams, B. Altshuler, Ya. Blanter, E. Fradkin, A. Larkin,
A. Millis, A. M. Sengupta, F. Sols, M. Turlakov, H. Westfahl and especially
I. Aleiner and A. J. Leggett were of much help to my understanding of the problem.
Kind hospitality of the Aspen Center for Physics during the 1999 workshop on
unconventional order in metals, where part of this study has been carried out,
is gratefully acknowledged. This work was supported by the MacArthur Chair endowed
by the John D. and Catherine T. MacArthur Foundation at the University of Illinois,
and by the National Science Foundation (DMR 91-20000) through the Science and
Technology Center for Superconductivity.


\begin{references}
\bibitem{ge} L. P. Gorkov, G. M. Eliashberg, 
Sov. Phys. JETP {\textbf{21}}, 940 (1965). 
\bibitem{e} K. B. Efetov, 
Sov. Phys. JETP {\textbf{55}}, 514 (1982). 
 \bibitem{as} B. L. Altshuler and B. I. Shklovskii, 
Sov. Phys. JETP {\textbf{64}}, 127 (1986). 
\bibitem{sa} 
B. D. Simons and B. L. Altshuler, in 
{\emph{Mesoscopic Quantum Physics}}, eds. E. Akkermans, 
G. Montambaux, J. -L. Pichard and J. Zinn-Justin  
(North Holland, Amsterdam, 1995); 
B. D. Simons and B. L. Altshuler, 
Phys. Rev. {\textbf{B 48}}, 5422 (1993). 
\bibitem{af} A. Altland and D. Fuchs, 
Phys. Rev. Lett. {\textbf{74}}, 4269 (1995). 
\bibitem{az2} A. Altland and M. Zirnbauer, 
Phys. Rev. Lett. {\textbf{76}}, 3420 (1996). 
\bibitem{asts} A. Altland, 
B. D. Simons, D. Taras-Semchuk, 
Advances in Physics {\textbf{49}}, 321 (2000). 
\bibitem{skf} M. A. Skvortsov, V. E. Kravtsov, 
M. V. Feigel'man, JETP Lett. {\textbf{68}}, 84 (1998). 
\bibitem{bcsz} R. Bundschuh, C. Cassanello, 
D. Serban, M. R. Zirnbauer, 
Nucl. Phys. {\textbf{B 532}}, 689 (1998).
\bibitem{kl} A. A. Koulakov, A. I. Larkin, 
Phys. Rev. {\textbf{B 60}}, 14597 (1999). 
\bibitem{i} D. A. Ivanov, preprint cond-mat/9911147. 
\bibitem{remark1} The constant \( C \) of the present work corresponds to \( ks^{2}/\beta  \) of Ref. \cite{as}, and \( k=\beta =1 \) .
\bibitem{m} M. L. Mehta, {\emph{Random Matrices}}, 
(Academic Press, 1991). 
\bibitem{az1} A. Altland and M. Zirnbauer, 
Phys. Rev. {\textbf{B 55}}, 1142 (1997).
 \bibitem{t} M. Tinkham, 
Phys. Rev. {\textbf{B 6}}, 1747 (1972). 
\bibitem{dg} P. G. de Gennes, 
{\emph{Superconductivity of Metals and Alloys}}, 
(Addison-Wesley, 1989), Ch. 5. 
\bibitem{sf} T. Senthil, 
M. P. A. Fisher, L. Balents and C. Nayak, 
Phys. Rev. Lett. {\textbf{81}}, 4704 (1998); 
T. Senthil and M. P. A. Fisher, 
Phys. Rev. {\textbf{B 60}}, 6893 (1999). 
\bibitem{s} See, e. g., J. R. Schrieffer, 
{\emph{Theory of Superconductivity}}, (Addison-Wesley, 1988), 
Ch. 8.
\bibitem{remark2} 
The Nambu notations break the full spin rotation 
symmetry, leaving intact only the rotations around \(\hat{z}\). 
\bibitem{agd} See, e. g., A. A. Abrikosov, L. P. Gorkov and I. E. Dzyaloshinski, 
{\emph{Methods of Quantum Field Theory in Statistical Physics}}, (Dover, 1975), 
Ch. 4.
\bibitem{remark3} The commonly used \cite{s} Ward 
identities relate \(\Sigma_{R}(\omega)\) with 
\(\langle \tau_{i} \rangle_{RR}\). 
The identities in the text are obtained in the 
Matsubara domain, by performing the analytic 
continuation with the two frequencies having 
the opposite sign of the imaginary part. 
\bibitem{remark4} Note that the statement that the level number variance \( \langle \delta N^{2}_{E}\rangle  \) in a d-wave grain is \( 3/4 \) of that in an s-wave grain, holds also in the diffusive limit \( E>E_{c} \), where \( \langle \delta N^{2}_{E}\rangle  \) depends on the sample dimensionality and is a function of \( E/E_{c} \).

\bibitem{remark5}However, I do not see how the spatial variation of the gap, which has to be taken into account at lower energies, would revive the charge diffusion mode. There is a possibility that the main result, Eq. (\ref{eq:answer}) with \( C=3 \) , is valid even beyond the formal applicability range of the calculation presented in this article. 
\bibitem{fe} A. L. Fetter, 
Phys. Rev. {\textbf{140}}, A1921 (1965). 
\end{references}
\end{document}